\documentclass[journal,twoside,web]{ieeecolor}
\usepackage{tmi}
\usepackage{cite}
\usepackage{amsmath,amssymb,amsfonts}
\usepackage{textcomp}
\usepackage[pdftex]{graphicx}
\usepackage[misc]{ifsym}
\usepackage[linesnumbered,ruled]{algorithm2e}
\usepackage{color, colortbl}
\usepackage{spverbatim}
\usepackage{etoolbox}
\usepackage{multirow}
\usepackage{booktabs}
\usepackage{courier}
\usepackage{bm}
\usepackage{bbm}
\usepackage{epstopdf}

\makeatletter\renewcommand\paragraph{\@startsection{paragraph}{4}{\z@}
  {.2em \@plus1ex \@minus.2ex}{-.5em}{\normalfont\normalsize\bfseries}}\makeatother
  
\newcommand{\norm}[1]{\left\lVert#1\right\rVert}
\def\BibTeX{{\rm B\kern-.05em{\sc i\kern-.025em b}\kern-.08em
    T\kern-.1667em\lower.7ex\hbox{E}\kern-.125emX}}
\begin{document}
\title{External Attention Assisted Multi-Phase Splenic Vascular Injury Segmentation with Limited Data}
\author{Yuyin~Zhou,
        David~Dreizin,
        Yan~Wang,
        Fengze~Liu,
        Wei~Shen,
        and~Alan~L.~Yuille, \IEEEmembership{Member, IEEE}
\thanks{This work was supported by University of Maryland Accelerated Translational Incubator Pilot (ATIP) Grant and NIH K08 EB027141-01A1.}
\thanks{Y. Zhou is with the Department of Computer Science and Engineering at University of California, Santa Cruz, CA 95064
    (e-mail: zhouyuyiner@gmail.com).}
\thanks{D. Dreizin is with the Dept of Radiology
R Adams Cowley Shock Trauma Center and University of Maryland, Baltimore, MD 21201, USA
    (e-mail: daviddreizin@gmail.com).}
\thanks{Y. Wang is with Shanghai Key Laboratory of Multidimensional Information Processing, East China Normal University, Shanghai 200241, China 
    (e-mail: wyanny.9@gmail.com).}
\thanks{F. Liu and A.L. Yuille are with the Department of Computer Science,
    the Johns Hopkins University, Baltimore, MD 21218, USA
    (e-mail: \{liufz13, alan.l.yuille\}@gmail.com).}
\thanks{W. Shen is with MoE Key Lab of Artificial Intelligence, AI Institute, Shanghai Jiao Tong University, Shanghai, China, 
    (e-mail: shenwei1231@gmail.com).}}

\maketitle

\begin{abstract}
The spleen is one of the most commonly injured solid organs in blunt abdominal trauma. 
The development of automatic segmentation systems from multi-phase CT for splenic vascular injury can augment severity grading for improving clinical decision support and outcome prediction.
However, accurate segmentation of splenic vascular injury is challenging for the following reasons: 
1) Splenic vascular injury can be highly variant in shape, texture, size, and overall appearance;
and 2) Data acquisition is a complex and expensive procedure that requires intensive efforts from both data scientists and radiologists, which makes large-scale well-annotated datasets hard to acquire in general.

In light of these challenges, we hereby design a novel framework for multi-phase splenic vascular injury segmentation, especially with limited data. 
On the one hand, we propose to leverage external data to mine pseudo splenic masks as the spatial attention, dubbed \emph{external attention}, for guiding the segmentation of splenic vascular injury.
On the other hand, we develop a \emph{synthetic phase augmentation} module, which builds upon generative adversarial networks, for populating the internal data by fully leveraging the relation between different phases. 
By jointly enforcing external attention and populating internal data representation during training, our proposed method outperforms other competing methods and substantially improves the popular DeepLab-v3+ baseline by more than 7\% in terms of average DSC, which confirms its effectiveness.
\end{abstract}


\section{Introduction}
\label{sec:introduction}
\begin{figure}[t]
\begin{center}
    \includegraphics[width=0.75\linewidth]{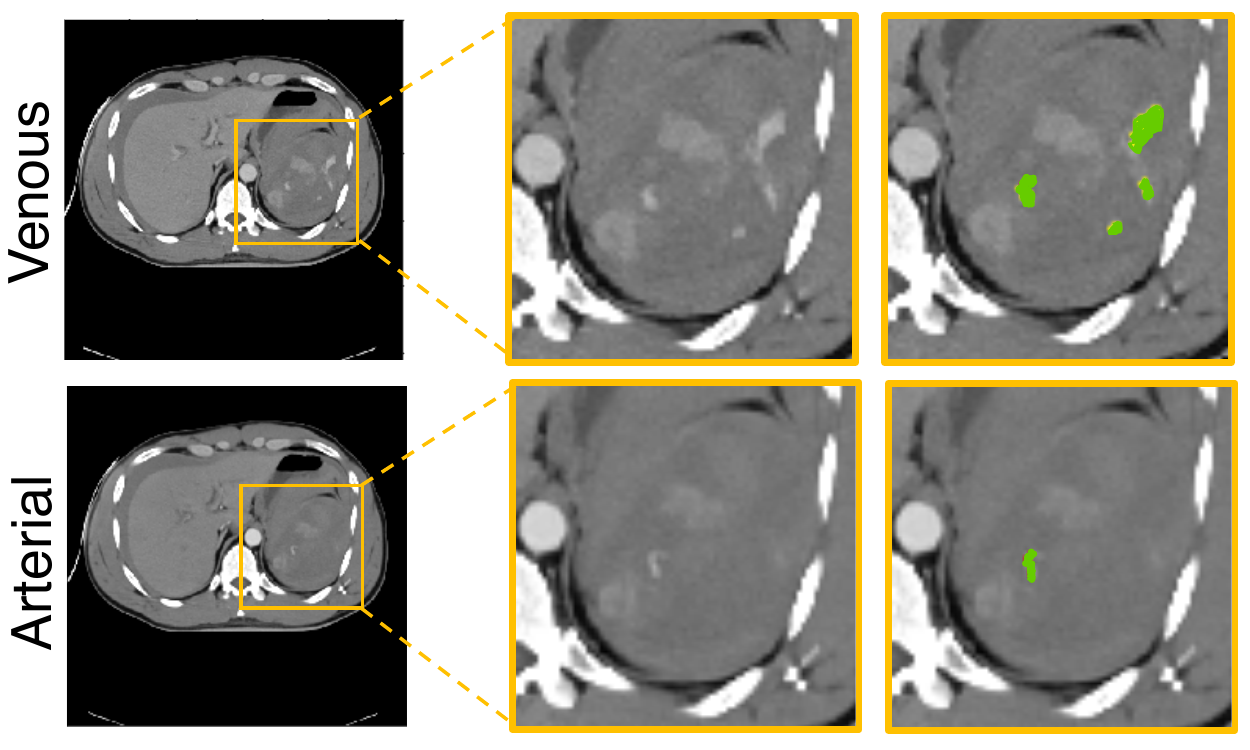}
\end{center}
\vspace{-1.5em}
\caption{Visual comparison of arterial and venous images (after alignment) as well as the manual segmentation of the splenic vascular injury. The injury volumes are different across phases.}
\vspace{-1.7em}
\label{Fig:Dataset}
\end{figure}

\IEEEPARstart{T}{he} spleen is the most commonly injured abdominal organ after blunt trauma~\cite{kozar2018organ,marmery2007optimization,bhangu2012meta}. 
Associated
splenic vascular injuries are risk factors for exsanguination, hemodynamic instability, and death.
CT is the first-line imaging modality to screen for splenic injuries in this setting~\cite{dreizin2012blunt}.
Vascular lesions include pseudoaneurysm and active bleeding. 
The degree of injury burden correlates with the need for interventions, namely angioembolization and splenectomy\cite{bhangu2012meta,smith2018management}. 
It is therefore important to be able to localize and quantify the volume of vascular injury on CT~\cite{champ2020}.
Contrast kinetics of contained pseudoaneurysms (PSAs) follow the blood pool, and PSAs are most conspicuous on arterial phase images. These
are typically either faint or imperceptible within the enhancing spleen on the portal phase. 
On the other hand, active arterial bleeding manifests as flame-shaped foci of hemorrhage that increase in size and decreases in density across phases~\cite{boscak2013optimizing}.
Automated quantification of splenic
vascular lesions would result in objective clinical imaging data for prognostication and
personalized decision support to predict mortality, transfusion requirement, and need for
intervention in this setting.
Therefore, in this paper, our goal is to automatically segment the splenic vascular injury from multi-phase (\emph{i.e.}, arterial and venous phase) images (examples illustrated in Fig.~\ref{Fig:Dataset}).

In recent years, deep learning has largely advanced the field of computer-aided diagnosis, especially in the field of biomedical image segmentation~\cite{dou2016automatic}\cite{Ronneberger_2015_UNet}\cite{Roth_2016_Spatial}\cite{zhu2019anatomynet}.
However, we have identified several limitations in the current literature.
Firstly, for challenging structures that show large inter-patient variation in terms of attributes such as shape, texture or size, even state-of-the-art architectures may yield less satisfactory performance in spite of their
good representational power~\cite{zhou2017fixed}.
Secondly, deep learning based approaches are data-driven and therefore require a large and representative training set which consists of medical images as well as their corresponding voxel-wise label maps.
However, in the cross-sectional medical imaging domain (particularly CT and MRI) where both images and annotations are expensive to acquire, rarely do we have such a perfectly-sized dataset to train a segmentation model~\cite{tajbakhsh2020embracing}. 
Data and labeling scarcity can be especially severe for applications such as splenic vascular injury segmentation, since collecting high-quality annotations from specially designated high-volume academic trauma centers can be even more challenging.
More critically, these two challenges can potentially add
to the difficulty of each other. 

To tackle these challenges, in this paper, we present a novel framework for multi-phase splenic vascular injury segmentation. 
Specifically, our method consists of: 1) an external attention assisted segmentation model, where we fully exploit external data for mining the spatial attention from the affiliated target (\emph{i.e.}, the spleen), referred to as \textbf{external attention}, for guiding the segmentation of splenic vascular injury;
and 2) an internal \textbf{synthetic phase augmentation} module, where we populate the internal data representation by leveraging the relation between arterial and venous phases.
By cooperatively enforcing external attention and augmenting the internal data representation, our method effectively addresses challenges arising from the large inter-patient variation of the splenic injury and data scarcity simultaneously.

In more concrete terms, the external attention assisted segmentation model aims to handle the variation of the discriminative features of splenic vascular injury, by allowing the network to focus on target structures of interest and neglect feature responses from more remote and irrelevant background regions.
To achieve this goal, existing approaches apply self-attention mechanisms (\emph{e.g.}, non-local operators) which exclusively rely on internal feature representations of the given target~\cite{wang2018non,schlemper2019attention}.
We hereby provide a novel perspective with the introduction of external attention, wherein the attention is extracted from external data with available supervision from the normal spleen segmentation to guide the relatively harder splenic vascular injury segmentation.
Mining external data has been known as a popular solution to address data scarcity~\cite{tajbakhsh2020embracing}.
This strategy has started to attract more and more research attention recently since it demonstrates great potential for boosting model performance and generalization~\cite{luo2020deep}. 
Here our goal is to fully exploit external data not only for expanding the size of the training data but rather to offer attentive knowledge for guiding the detection of splenic vascular lesions.

Inspired by the high relevance between
the location of a spleen and the associated injured regions (a splenic vascular injury resides either within or near the spleen), we formulate the unknown spleen\footnote{unknown spleen throughout this paper refers to the whole spleen on the internal splenic injury dataset, which is unknown due to that
the spleen class is not labeled.} location as a latent variable for deriving the attention to facilitate the following training process. We note the integration of our internal splenic vascular injury dataset and the external dataset yields a partially-supervised setting, in that each set only has either the spleen or the injury labeled, but not both.
An intuitive approach is to impute the missing category with pseudo-labels~\cite{lee2013pseudo} and integrate both sets in a joint training paradigm.
Nonetheless, this approach only addresses the label discrepancy between the internal and the external set without considering the correlation between different classes, \emph{e.g.}, a vascular injury (one class) often resides close to and emanate from the associated spleen (another class).
To fully exploit the spatial relationship between the spleen and its associated injury, we re-formulate the predicted splenic mask not only as pseudo-supervision but also to act as the spatial attention mask to reweight the importance of different voxels during training.
Specifically, during the learning process, the pseudo-labels on the unknown spleen locations and the segmentation network parameters are alternately updated, making the derived attention mask iteratively refined to facilitate the subsequent training iterations.
This allows the network to gradually focus exclusively on relevant foreground regions, so as to largely benefit the segmentation for the splenic vascular injury ~\cite{zhou2017fixed}.

To further address the limited training data, especially in the context of multi-phase data,  we develop a synthetic phase augmentation module by exploiting the underlying relationship between different imaging phases.
We use CycleGAN~\cite{zhu2017unpaired}, to build a phase translation model, and use the learned transformations to generate synthetic phases. 
The real and the generated multi-phase images are then jointly trained using our proposed external attention assisted segmentation model.
Consequently, the augmented synthetic examples not only greatly enrich the training set but also benefit the knowledge integration from both phases. 

In this paper, we have curated a patient cohort of 55 consecutive multi-phase CT studies with splenic vascular injury annotations, which, to the best of our knowledge, is the largest available set to date for this injury type. By evaluating this dataset comprehensively, our approach can consistently outperform prior arts by a large margin. The main contributions of this paper are summarized as follows:
\begin{itemize}
    \item We present a novel approach for multi-phase splenic vascular injury segmentation, which is a common and clinically important entity, yet has been rarely studied.
    \item We establish a new attention mechanism, where the attention is explicitly extracted from external data with available supervision for associated affiliated targets, referred to as \emph{external attention}.  
    \item We introduce \emph{synthetic phase augmentation}, where synthetic phases are created to be jointly trained with real phases to further benefit the knowledge integration from limited training data. 
    \item Extensive experiments conducted under various settings suggest that our approach not only significantly outperforms other competing methods for splenic vascular injury segmentation, but also well generalizes to liver tumor segmentation and pancreatic tumor segmentation.
\end{itemize}

\section{related works}
\label{sec:relatedworks}
\subsection{Training with External Data}
To enlarge the size of the training data and populate the training distribution, one popular strategy is to employ external unlabeled data or heterogeneous labeled datasets. This requires no additional manual efforts but can generally lead to imperfect training data~\cite{tajbakhsh2020embracing}. 

In terms of unlabeled data, Bai \emph{et al.}~\cite{bai2017semi} present a self-training-based method for cardiac MR 
image segmentation, where the network parameters and the pseudo-labels were alternatively updated. 
Zhou \emph{et al.}\cite{zhou2019semi,xia2020uncertainty} further propose to co-train multiple networks where
the pseudo-labels are refined by exploiting the consensus of network predictions in the ensemble. 
To make the learned models more robust, consistency-based methods~\cite{li2020transformation,liu2020semi} and uncertainty-driven approaches~\cite{yu2019uncertainty,wang2020ud} are proposed for different medical image classification and segmentation tasks.

For incorporating external labeled datasets, Lenga \emph{et al.}~\cite{lenga2020continual} demonstrate that joint training on multiple heterogeneous datasets leads to superior performance for chest X-ray classification.
To properly handle the variation of image characteristics and labeling protocols among different datasets, strategies for bridging domain gaps, imputing missing annotations, regularizing and stabilizing predictions are proposed for obtaining a universal feature representation~\cite{huang20193d,luo2020deep,dmitriev2019learning,guendel2018learning,fang2020multi,huang2020multi,shi2020marginal,tang2020learning}.

However, in this paper, we provide a different perspective, that is, we propose to utilize external data to incorporate attention for learning the affiliated target. 

\vspace{-1.2em}
\subsection{Multi-phase Learning}
Medical images acquired in different phases usually contain complementary information, hence radiologists usually rely on analyzing multi-phase data for better image interpretations.
Recently, researchers have explored different strategies for combining multi-phase images, such as joint training~\cite{yan2020learning,raju2020co,huo2020harvesting,zhang2020robust}, feature-level fusion~\cite{zhou2019hyper,xia2020detecting}, and generative models~\cite{zheng2018phase}. In this paper, similar to~\cite{zheng2018phase}, we also use generative models to create synthetic multi-phase images which can be then used for training the segmentation model.

\vspace{-1.1em}
\subsection{Synthetic Data Augmentation}
To alleviate overfitting and improve model performance, researchers propose to use Generative Adversarial Networks (GANs) for synthetic data augmentation.
Many studies such as~\cite{frid2018synthetic,sandfort2019data,bowles2018gan} have suggested that generative adversarial networks generate convincing appearances of CT. For instance, Frid-Adar~\emph{et al.}~\cite{frid2018synthetic} show that the synthetic lesions by GANs are meaningful in appearance, and using the generated examples as augmentation achieves a significant improvement of 7\% for liver lesion classification. Sandfort~\emph{et al.}~\cite{sandfort2019data} demonstrate that the synthetic non-contrast images (using CycleGAN, a variant of GANs) appear convincing - even when significant abnormalities are present in the contrast CT scans. 
Different from these methods, here we use CycleGAN to exploit the multi-phase information 
and generate synthetic imaging phases to benefit the segmentation of multi-phase splenic injury.
The generated images are augmented in the training set to benefit the knowledge integration from multi-phase images, especially from a limited set.

\vspace{-1.1em}
\subsection{Attention Mechanisms}
The attention mechanism has been
widely applied to many vision problems. 
Wang \emph{et al.}~\cite{wang2018non} propose to model long-range relationships and design a non-local operator accordingly.
In the field of medical image analysis, Zhou \emph{et al.}~\cite{zhou2017fixed,yu2018recurrent} propose a multi-stage framework where the first stage explicitly extracts the saliency region, which is then used for facilitating the segmentation in the second stage for better detection of small organs such as the pancreas.  
To further reduce the computational resources and model parameters, Schlemper~\emph{et al.}~\cite{schlemper2019attention} propose additive attention
gate modules which are integrated in the skip connections.
Additionally, attention modules are proposed to be used at multiple resolutions which are then fused for prostate segmentation~\cite{wang2019deep}.

Different from the works above, we propose to incorporate attention for the splenic vascular injury segmentation by leveraging the spatial relevance between the spleen and the injury.
Therefore, the attention in our study is derived by mining the unknown spleen class with the help of additional external data.

\begin{figure*}[t]
\begin{center}
    \includegraphics[width=0.95\linewidth]{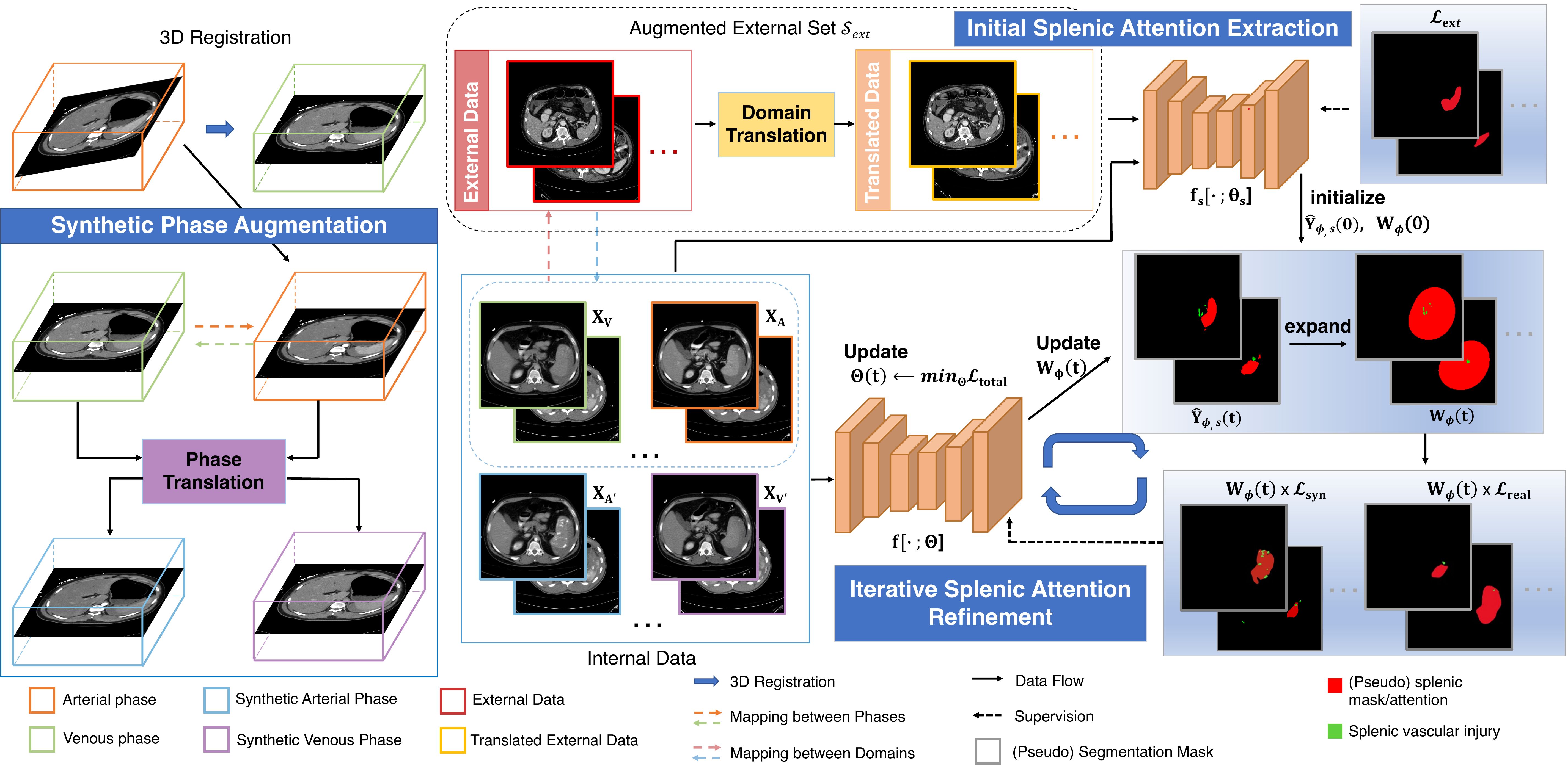}
\end{center}
\vspace{-1.5em}
\caption{The overall framework. We first train a spleen segmentation model $\mathbf{f}_s\!\left[\cdot;\boldsymbol{\theta}_s\right]$ on the external set $\mathcal{S}_{ext}$. 
For a given input $\mathbf{X}_\phi$, the corresponding pseudo splenic mask $\hat{\mathbf{Y}}_{\phi,s}$ and the attention map $\mathbf{W}_\phi$ are first initialized by inferring on $\mathbf{f}_s\!\left[\cdot;\boldsymbol{\theta}_s\right]$. A domain translation model is also trained for generating synthetic external data as augmentation.
Secondly, a phase translation model is trained for creating synthetic arterial and venous phases.
The real and synthetic phases are then jointly trained with the attention map applied on top of the losses as spatial weights, during which the attention $\mathbf{W}_\phi$ can be iteratively refined.
We can see that the attention mask can well enclose the injured regions by expanding. 
} 
\label{Fig:framework}
\vspace{-1.5em}
\end{figure*}

\section{Methodology}

\subsection{Overview}
In the context of multi-phase splenic vascular injury segmentation,
the data and the associated annotations can be categorized into two sets based on the imaging phases: $\mathcal{A}=\{\mathbf{X}_\mathrm{A}\in \mathbb{R}^{D_\mathrm{A}},\mathbf{Y}_\mathrm{A}\in\mathbbm{L}_{int}^{D_\mathrm{A}}\}$ and $\mathcal{V}=\{\mathbf{X}_\mathrm{
V}\in \mathbb{R}^{D_\mathrm{
V}},\mathbf{Y}_\mathrm{V}\in\mathbbm{L}_{int}^{D_\mathrm{
V}}\}$, where $\mathbf{X}_\mathrm{A}$, $\mathbf{X}_\mathrm{V}$ and $\mathbf{Y}_\mathrm{A}$, $\mathbf{Y}_\mathrm{V}$ denote the image and the annotation of dimension $D_\mathrm{A}$, $D_\mathrm{V}$ respectively; the subscripts A and V, denote \emph{arterial} and \emph{venous} phases in CT imaging protocols.  Note that unlike previous studies~\cite{Havaei_2017_Brain,Kamnitsas_2016_Efficient,zhou2019hyper}, here the dataset is not paired, \emph{i.e.}, the arterial phase and the venous phase of the same patient do not have identical annotations ($\mathbf{Y}_\mathrm{A}\neq\mathbf{Y}_\mathrm{V}$).
$\mathbbm{L}_{int}=\{\textup{background}, \textup{splenic vascular injury}\}$ denotes the labeling space of the internal multi-phase splenic vascular injury dataset.
Additionally, we also introduce an external dataset $\mathcal{S}_{ext}=\{\mathbf{X}_{ext}\in \mathbb{R}^{D_\mathrm{E}},\mathbf{Y}_{ext}\in {\mathbbm{L}_{ext}^{D_\mathrm{E}}}\}$ where the spleen supervision is available,~\emph{i.e.}, $\mathbbm{L}_{ext}= \{\textup{background}, \textup{spleen}\}$. Image $\mathbf{X}_{ext}$ and annotation $\mathbf{Y}_{ext}$ are of the same dimension $D_\mathrm{E}$. 
Note that the external dataset only contains normal abdominal CT scans, \emph{i.e.}, there is no splenic vascular injury available, whereas in the internal dataset, only voxel-wise splenic vascular injury annotations are given. 
Therefore, our goal of using both datasets to boost splenic vascular injury segmentation also belongs to the partially-supervised setting.

Fig.~\ref{Fig:framework} illustrates the overall pipeline, which can be summarized into the following two stages.
In the first stage, we train a spleen segmentation model $\mathbf{f}_s\!\left[\cdot;\boldsymbol{\theta}_s\right]$ exclusively on the external set $\mathcal{S}_{ext}$, in order to obtain a good initialization of the external attention (Initial  Splenic  Attention  Extraction). 
For a given input $\mathbf{X}_\phi$, the corresponding pseudo splenic mask $\hat{\mathbf{Y}}_{\phi,s}$ and the attention map $\mathbf{W}_\phi$ (of the same size as $\mathbf{X}_\phi$) are first initialized by inferring on the trained model and later updated in the following stage.
Here we also train a domain translation model to generate synthetic external data as augmentation.
In the second stage, a phase translation model is trained to exploit the relationship between different phases, and the learned mapping functions are used to create synthetic arterial and venous phases (Synthetic Phase Augmentation).
The real and synthetic phases are then jointly trained with the attention map applied on top of the loss function as spatial weights, where the network $\mathbf{f}\!\left[\cdot;\boldsymbol{\Theta}\right]$, the pseudo splenic mask $\hat{\mathbf{Y}}_{\phi,s}$, and the attention map $\mathbf{W}_\phi$ are alternately updated (Iterative Splenic Attention Refinement).
Below, we will elaborate on each component along with technical details.

\subsection{Initial Splenic Attention Extraction}
\label{sec:init_spln_att}

The spleen is a small organ compared to the whole abdominal region. In comparison, the splenic vascular injury is even smaller.
In our internal dataset, the fraction of the splenic vascular injury, relative to the entire
volume, is often much smaller than 0.005\%. 
This largely increases the difficulty of direct segmentation or even localization from the entire CT volume, which coincides with previous findings for pancreatic cyst segmentation~\cite{Zhou_2017_Deep}.
To deal with this problem, Zhou \emph{et al.}~\cite{Zhou_2017_Deep} suggest that starting from the pancreas mask can largely increase
the chance of accurately segmenting the cyst and propose to perform cyst segmentation based on the pancreatic region,
which is relatively easy to detect. However, directly applying this method would require voxel-wise annotations for the spleen, which are not available in our internal dataset. 
Instead, we have sought to $\mathcal{S}_{ext}$ for effective splenic attention extraction to guide the splenic vascular injury segmentation in the following stage (Sec.~\ref{sec:iter_att}). 
To bridge the gap between internal and external domains, we also train a \textbf{domain translation model} and use the translated data as augmentation.
In other words, the external set $\mathcal{S}_{ext}$ used for training the initial attention extraction model here is the union of both the external data and the translated data (see Fig.~\ref{Fig:framework}).
The translated data are generated from the external data by training a domain translation model which learns the mapping between the internal and the external domains. The technical details for training the domain translation model are essentially the same as used in the phase translation model (see Sec. III-C).

To obtain a good feature extractor for the spleen, we first optimize the cross-entropy loss exclusively on the external set as follows:
\begin{equation}\label{cross_entropy}
\begin{split}
    -\sum_{l\in\mathbbm{L}_{ext}}\sum_{j} &\mathbbm{1}(y_{ext,l}^j)\log p_{ext,l}^j,
\end{split}
\end{equation}
\noindent where $p_{ext,l}^j$ denotes the probability of class $l$ on the $j$-th voxel.  

To facilitate the detection of the splenic vascular injury in the later stage (Sec.~\ref{sec:iter_att}), for a given input $\mathbf{X}_\phi$ ($\phi$ is the imaging phase), we first compute the prediction for the unknown spleen class as the initial pseudo splenic mask $\hat{\mathbf{Y}}_{\phi,s}^{(0)}$.
Considering that the splenic vascular injury can sit outside the spleen, we further relax the predicted boundary to compute the initial splenic attention map $\mathbf{W}_\phi^{(0)}$ via a transformation function $\boldsymbol{r}(\cdot)$:
\begin{equation}\label{eqn:init_attention}
\begin{split}
\mathbf{W}_\phi^{(0)}=\boldsymbol{r}(\hat{\mathbf{Y}}^{(0)}_{\phi,s}, \sigma)={\mathbb{I}\!\left[(\hat{\mathbf{Y}}^{(0)}_{\phi,s} \star \boldsymbol{N}(\sigma))\geqslant\rho\right]},
\end{split}
\end{equation}
where $\boldsymbol{N}(\sigma)$ is the Gaussian kernel with the standard deviation of $\sigma$. $\star$ denotes the convolution operator and $\rho$ is the threshold for generating the attention mask. $\mathbb{I}(\cdot)$ is the indicator function.
As can be seen from Fig.~\ref{Fig:framework}, the transformed attention mask can successfully enclose exterior injuries through expanding.

\subsection{Synthetic Phase Augmentation via Generative Adversarial Networks}
\label{sec:multi-phase}
Due to the different imaging protocols, $\mathcal{A}$ and $\mathcal{V}$ demonstrate two distributions of different appearances. Meanwhile, they are also highly correlative since different imaging phases are still corresponding to the same patient. 
Thereby, how to incorporate multi-phase information to boost model performance has become a promising direction to explore~\cite{zhou2019hyper,zheng2018phase,yan2020learning,raju2020co,huo2020harvesting,zhang2020robust}.
The goal is thus to train a segmentation model for both $\mathcal{A}$ and $\mathcal{V}$ simultaneously, so that the information in these two phases are considered in a collaborative way. 

Following~\cite{zheng2018phase}, we use the CycleGAN model~\cite{zhu2017unpaired} which aims to learn mapping functions between domains, to build relation between the two phases, referred to as the phase translation model.  
Our goal here is to learn mapping functions $\boldsymbol{g}:\mathbf{X}_\mathrm{A}\rightarrow \mathbf{X}_\mathrm{V}$ and $\boldsymbol{h}:\mathbf{X}_\mathrm{V}\rightarrow \mathbf{X}_\mathrm{A}$. 
Before training the phase translation model, we first perform \textbf{3D registration} where all arterial phase CT scans of the same patient are pre-registered
to the venous phase using DEEDS~\cite{heinrich2013mrf}.
The goal of the mapping function $\boldsymbol{g}(\cdot)$ is to generate images $\boldsymbol{g}(\mathbf{X}_\mathrm{A})$ indistinguishable from $\mathbf{X}_\mathrm{V}$ and minimise the loss function in Eqn.~(\ref{eqn:A2V}), where the discriminator $\boldsymbol{D}_\mathrm{V}(\cdot)$ is
trying to distinguish the generator’s generated images and maximize the overall loss:
\begin{align}
    \mathrm{min}_{\boldsymbol{g}} \mathrm{max}_{\boldsymbol{D}_\mathrm{V}}\mathcal{L}_{\mathrm{A}\rightarrow\mathrm{V}} =& \ \mathrm{min}_{\boldsymbol{g}} \mathrm{max}_{\boldsymbol{D}_\mathrm{V}}(\mathbb{E}_{\mathbf{X}_\mathrm{V} \sim \mathcal{V}}[\log \boldsymbol{D}_\mathrm{V}(\mathbf{X}_\mathrm{V})] \nonumber \\
  +& \ \mathbb{E}_{\mathbf{X}_\mathrm{A} \sim \mathcal{A}}[\log (1-\boldsymbol{D}_\mathrm{V}(\boldsymbol{g}(\mathbf{X}_\mathrm{A}))]),
\label{eqn:A2V}
\end{align}

To learn the mapping function $\boldsymbol{h}(\cdot)$, a similar adversarial loss \emph{w.r.t.} $\boldsymbol{h}(\cdot)$ and its discriminator $\boldsymbol{D}_\mathrm{A}(\cdot)$ is optimized in the same manner, \emph{i.e.}, $\mathop{\mathrm{min}}_{\boldsymbol{h}} \mathop{\mathrm{max}}_{\boldsymbol{D}_\mathrm{A}}\mathcal{L}_{\mathrm{V}\rightarrow\mathrm{A}}$. Additionally, a cycle-consistency loss is imposed:
\begin{align}
    \mathcal{L}_{\text{cyc}} =  & \ \mathbb{E}_{\mathbf{X}_\mathrm{A} \sim \mathcal{A}}[\norm{\boldsymbol{h}(\boldsymbol{g}(\mathbf{X}_\mathrm{A}))-\mathbf{X}_\mathrm{A}}_1] \nonumber \\ 
    + &\ \mathbb{E}_{\mathbf{X}_\mathrm{V} \sim \mathcal{V}}[\norm{\boldsymbol{g}(\boldsymbol{h}(\mathbf{X}_\mathrm{V}))-\mathbf{X}_\mathrm{V}}_1].
\end{align}
Therefore, the final objective function for training the phase translation model $\boldsymbol{g(\cdot)}$ and $\boldsymbol{h(\cdot)}$ is:
\begin{equation}
\label{Eqn:translation_losses}
\begin{split}
    \mathcal{L}_{\mathrm{A}\rightarrow\mathrm{V}} + \mathcal{L}_{\mathrm{V}\rightarrow\mathrm{A}} +
    \lambda  \mathcal{L}_{\text{cyc}}, 
\end{split}
\end{equation}   
where $\lambda$ is the hyper-parameter for balancing the loss terms.

\begin{figure}[t]
\begin{center}
    \includegraphics[width=0.95\linewidth]{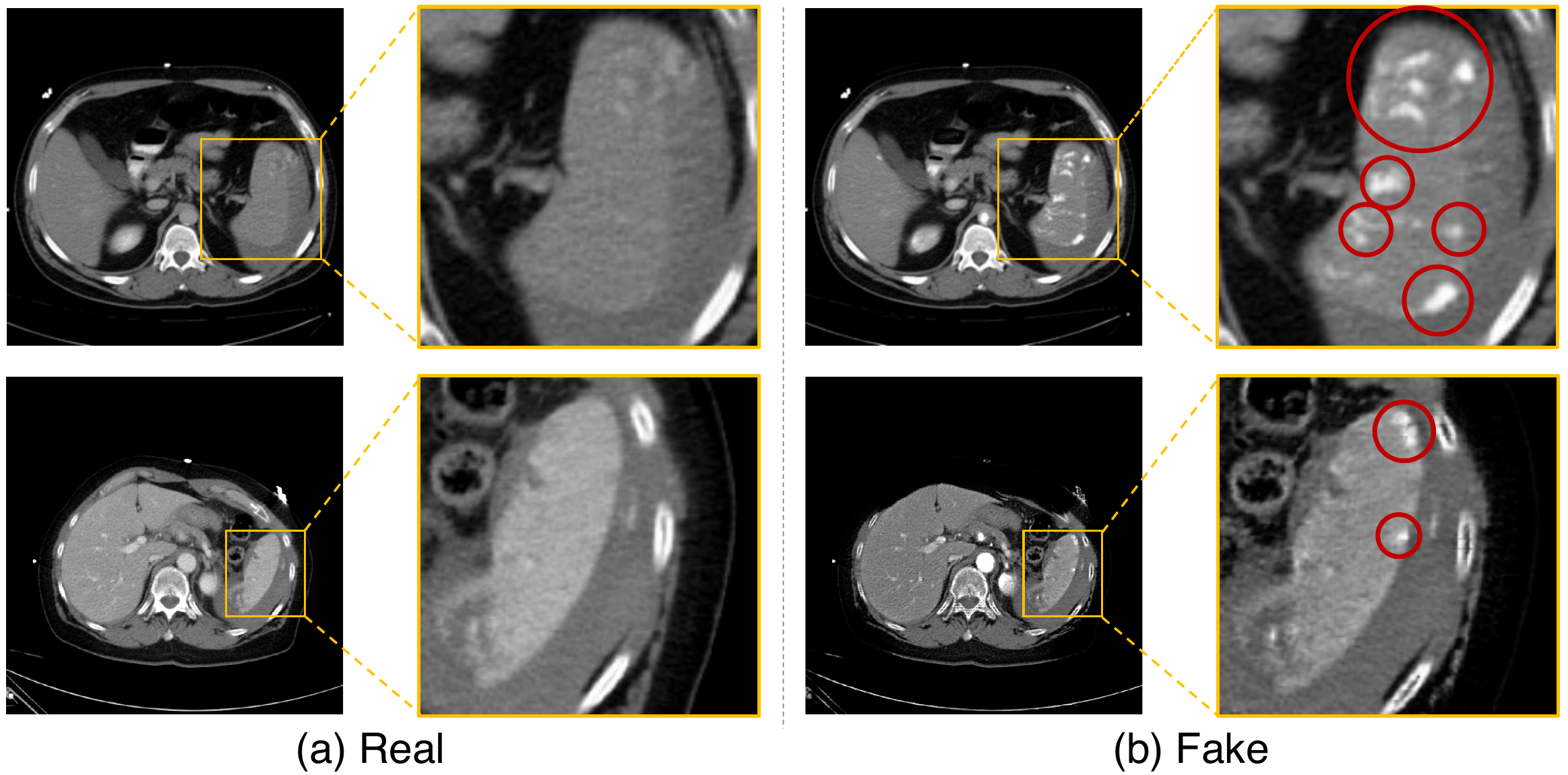}
\end{center}
\vspace{-1.5em}
\caption{Real images vs. synthetic images. Synthesized injury spots are inpainted onto the generated images.} 
\vspace{-1.5em}
\label{Fig:gan_example}
\end{figure}

Fig.~\ref{Fig:gan_example} illustrates a few pairs of examples of the original image and the generated synthetic image.    
We can see that our phase translation model have successfully brought contrast and texture changes to the original images, which in turn make them appear more similar to images from the other phase.    
This means that the new texture of the splenic vascular injury can be inpainted onto the generated images. 
Therefore, unlike previous studies~\cite{sandfort2019data,bowles2018gan}, here the real and the synthetic data will be jointly trained using the subsequent segmentation network in a self-learning manner.
Self-learning, or self-training, usually consists the following steps: 1) train a teacher model on labeled images and use it to propagate labels from the labeled to the unlabeled data; 2) then use the larger set combined of both labeled and pseudo labeled data for training a student model. Then this process can be iterated by putting back the student as the teacher. This approach has been successfully applied to multiple natural and medical imaging tasks~\cite{xie2020self,zhang2018self,lee2013pseudo,bai2017semi,zhou2019semi}.
In our method, similarly, we train the external attention assisted model in a similar manner where it propagates labels from the internal labeled images to the synthetic images, and then using the larger, newly labeled set for training. The detailed training algorithm is illustrated next.

\subsection{Iterative Attention Refinement via Unknown Class Mining}
\label{sec:iter_att}
Motivated by the high relevance between a spleen and its associated injury, we propose to formulate the splenic mask as spatial attention to guide the learning process of the splenic vascular injury.
Unlike previous attention-based methods~\cite{schlemper2019attention}, here we aim to mine additional information from the unknown spleen category by incorporating external data with available splenic supervision.
An intuitive approach is to 
leverage the power of pseudo-labels~\cite{lee2013pseudo}.
However, here we provide a new perspective---to exploit the predicted splenic mask not only as pseudo-supervision in the training process but also to provide attention to let the network focus exclusively on relevant foreground regions. 
With a proper transformation function to account for exterior injuries (Sec.~\ref{sec:init_spln_att}), an attention mask can be computed from the pseudo splenic mask to reweight the importance of different voxels.
Since the attention is originated from external data, we refer to our model as external attention assisted segmentation model, denoted by $\mathbf{f}[\cdot;\boldsymbol{\Theta}]$.
As the pseudo-labels of the unknown voxels and the network parameters $\boldsymbol{\Theta}$ are alternately updated, the derived external attention can be iteratively refined in the learning process.
Specifically at the $t$-th update, by applying the attention mask $\mathbf{W}_\phi^{(t)}$ as spatial weights to the internal loss terms computed on the splenic vascular injury dataset,
the loss on real arterial and venous phases ($\mathrm{A}/\mathrm{V}$) can be written as:
\begin{align}
\label{Eqn:loss_real}
\begin{split}
{\mathcal{L}_{real}}= 
-\sum_{\phi\in\{\mathrm{A}, \mathrm{V}\}}&\sum_{j}w_\phi^j\cdot[
\mathbbm{1}(\hat{y}_{\phi, s}^j)\log p_{\phi, s}^j + \\& \mathbbm{1}(y_{\phi, i}^j)\log p_{\phi, i}^j + \mathbbm{1}(\hat{y}_{\phi, b}^j)\log p_{\phi, b}^j], 
\end{split}
\end{align}
where the subscripts $s,i,b$ stand for the spleen, injury and background class, respectively.
$p_\phi^j$ is the $j$-th output of the computed probability map $\mathbf{f}\!\left[\mathbf{X}_\phi;\boldsymbol{\Theta}\right]$. 
$y^j_{\phi, i}$ denotes the $j$-th voxel of the splenic injury manual segmentation mask $\mathbf{Y}_{\phi, i}$. 
For real data where the injury annotation is given, we need to re-estimate the pseudo-label of the spleen class ($\hat{\mathbf{Y}}_{\phi, s}$) and the background class ($\hat{\mathbf{Y}}_{\phi, b}$).
Therefore for non-injury regions, $\hat{y}^j_{\phi, s} = 1, \hat{y}^j_{\phi, b} = 0$ if $p^j_{\phi, s} > p^j_{\phi, b}$, otherwise $\hat{y}^j_{\phi, s} = 0, \hat{y}^j_{\phi, b} = 1$.
$w_\phi^j$ is the $j$-th voxel of the attention mask $\mathbf{W}_\phi^{(t)}$.

\begin{algorithm}[t!]
\SetKwInOut{Input}{Input}
\SetKwInOut{Output}{Output}
\SetKwInOut{Return}{Return}
\Input{
    Internal image $\mathbf{X}_\phi$  (${\phi=\{\mathrm{A,V,A',V'}}\}$) and labelmap $\mathbf{Y}_{\phi,i}$; \\
    ~External image $\mathbf{X}_{ext}$ and labelmap $\mathbf{Y}_{ext,s}$;\\
    ~Max number of iterations $T$;\\
    ~Hyper-parameters $\alpha, \beta$;
}
\Output{
    Network $\mathbf{f}\!\left[\cdot;\boldsymbol{\Theta}\right]$;
}
Training a spleen segmentation model $\mathbf{f}_s\!\left[\cdot;\boldsymbol{\theta}_s\right]$  on the external set;\\
Extract initial splenic attention as $\mathbf{W}_\phi^{(0)}=\boldsymbol{r}(\hat{\mathbf{Y}}^{(0)}_{\phi,s}, \sigma)$, 
where $\hat{\mathbf{Y}}^{(0)}_\phi=\mathrm{arg\,max}_{l\in \mathbb{L}_{ext}}\mathbf{f}_s[\mathbf{X}_\phi;\boldsymbol{\theta}_s]$;\\
${t}\leftarrow{0}$; \\
\Repeat{${t}={T}$}{
    ${t}\leftarrow{t+1}$;\\
    Fix $\mathbf{W}_\phi=\mathbf{W}^{(t-1)}_\phi$ and $\hat{\mathbf{Y}}_{\phi,s}=\hat{\mathbf{Y}}^{(t-1)}_{\phi,s}$: \\
    ~~~~Compute $\mathcal{L}_{real}$ and $\mathcal{L}_{syn}$ on the internal set by Eqn.~\eqref{Eqn:loss_real} \& Eqn.~\eqref{Eqn:loss_syn}; \\
    ~~~~Compute $\mathcal{L}_{ext}$ on the external set by Eqn.~\eqref{Eqn:loss_ext};\\
    ~~~~Compute $\mathcal{L}_{total}$ by Eqn.~\eqref{Eqn:total_seg_loss};\\
    
    ~~~~Update the segmentation model ${\mathbf{f}[\mathbf{X}_\phi;\boldsymbol{\Theta}^{(t)}]}$ by
    $\boldsymbol{\Theta}^{(t)}\leftarrow\mathop{\mathrm{min}}_{\boldsymbol{\Theta}}{\mathcal{L}_{total}}$;\\
    
    ${\hat{\mathbf{Y}}^{(t)}_\phi}\leftarrow\mathrm{arg\,max}_{l \in \mathbb{L}_{int} \cup \mathbb{L}_{ext}}\,    {\mathbf{f}\!\left[\mathbf{X}_\phi;\boldsymbol{\Theta}^{(t)}\right]}$; \\
    $\mathbf{W}_\phi^{(t)}=\boldsymbol{r}(\hat{\mathbf{Y}}^{(t)}_{\phi,s}, \sigma)$;\\
}
\Return{
    $\boldsymbol{\Theta}=\boldsymbol{\Theta}^{(T)}$.
}
\caption{
    External Attention Assisted Training
}
\label{Alg:Training}
\end{algorithm}

Similarly, the loss on synthetic arterial and venous phases ($\mathrm{A'}/\mathrm{V'}$) can be written as:
\begin{align}
\label{Eqn:loss_syn}
\begin{split}
{\mathcal{L}_{syn}}= 
-\sum_{\phi\in\{\mathrm{A'}, \mathrm{V'}\}}&\sum_{j}w_\phi^j\cdot[\mathbbm{1}(\hat{y}_{\phi, s}^j)\log p_{\phi, s}^j + \\& \mathbbm{1}(\hat{y}_{\phi, i}^j)\log p_{\phi, i}^j + \mathbbm{1}(\hat{y}_{\phi, b}^j)\log p_{\phi, b}^j], 
\end{split}
\end{align}
where $\hat{y}^j_{\phi, s}$, $\hat{y}^j_{\phi, i}$ denotes the $j$-th label of the pseudo splenic mask $\hat{\mathbf{Y}}_{\phi, s}$ and pseudo splenic injury mask $\hat{\mathbf{Y}}_{\phi, i}$ respectively.
Consequently, the $j$-th label of the backgound class can be computed as $\hat{y}^j_{\phi, b} = 1 - \hat{y}^j_{\phi, s} - \hat{y}^j_{\phi, i}$. 
Unlike typical synthetic data augmentation where the augmented training samples are often assumed to have the same labelmaps as the original data, here we use the pseudo splenic injury mask $\hat{\mathbf{Y}}_{\phi, i}$ for computing the loss on the generated phases,
following~\cite{bai2017semi,zhou2019semi}.
This is due to that synthetic injuries can be inpainted onto generated images, making the original labelmaps no longer applicable for injury segmentation on these generated images (Sec.~\ref{sec:multi-phase}, Fig.~\ref{Fig:gan_example}).
In addition, the loss on the external set can be computed as follows:
\begin{align}
\label{Eqn:loss_ext}
\begin{split}
&{\mathcal{L}_{ext}}= 
-\sum_{j} [\mathbbm{1}(y_{ext, s}^j)\log p_{ext, s}^j + \mathbbm{1}(y_{ext, b}^j)\log p_{ext, b}^j], 
\end{split}
\end{align}
where $p_{ext}^j$ is the $j$-th output of the computed probability map $\mathbf{f}\!\left[\mathbf{X}_{ext};\boldsymbol{\Theta}\right]$. 
Therefore, the overall loss function becomes a weighted sum of both internal and external losses:
\begin{align}
\label{Eqn:total_seg_loss}
\begin{split}
&{\mathcal{L}_{total}}=\alpha \mathcal{L}_{real}+ (1-\alpha) \mathcal{L}_{syn} +\beta\mathcal{L}_{ext},
\end{split}
\end{align}
\noindent where we introduce a coefficient $\alpha \in [0.0, 1.0]$ to adjust the weight of the original and synthetic data.
Specially, $\alpha = 1.0$ indicates that synthetic phases are not used in the training.
$\beta$ is used for balancing the weight between the internal and external data.

During the learning process, the attention mask $\mathbf{W}_\phi$ is  initialized by Eqn.~\eqref{eqn:init_attention}, and then iteratively refined during the training process.
After the $t$-th update of the pseudo-labels, the attention mask $\mathbf{W}_\phi^{(t)}$ is then expanded (Sec.~\ref{sec:init_spln_att}) for reweighting the loss function to further facilitate the following training process to learn $\boldsymbol{\Theta}^{(t+1)}$ (Eqn.~\eqref{Eqn:loss_real} \& Eqn.~\eqref{Eqn:loss_syn}). $\hat{\mathbf{Y}}_\phi^{\left(t\right)}$ and $\mathbf{W}_\phi^{(t)}$ can be written as:
\begin{eqnarray}
\label{Eqn:pseudo_gt_update}
{\hat{\mathbf{Y}}_\phi^{\left(t\right)}}&=&\mathrm{arg\,max}_{l \in \mathbb{L}_{int} \cup \mathbb{L}_{ext}}\,    {\mathbf{f}\!\left[\mathbf{X}_\phi;\boldsymbol{\Theta}^{(t)}\right]}, \\
\label{Eqn:attention_update}
{\mathbf{W}_\phi^{(t)}}&=&\boldsymbol{r}({\hat{\mathbf{Y}}^{\left(t\right)}_{\phi, s}, \sigma}),
\end{eqnarray}
where ${t}={0,\ldots,T-1}$ is the index of iteration. 
As indicated by Eqn.~\eqref{Eqn:pseudo_gt_update}~\&~\eqref{Eqn:attention_update}, in our framework, the attention map can be iteratively refined by exploiting the unknown spleen class.
The overall training procedure is illustrated in Algorithm~\ref{Alg:Training}. It consists of two training stages: 1) \textbf{Initial Splenic Attention Extraction}, where the initial attention is extracted by training a spleen segmentation model $\boldsymbol{\theta}_s$ on \textbf{the external data};
2) \textbf{Iterative Attention Refinement}, where the attention mask is initialized by the initial splenic attention mask. Then we train another segmentation network $\boldsymbol{\Theta}$ on \textbf{the internal dataset} with the attention map  applied  on  top  of  the  loss  function  as  spatial  weights. The network parameter $\boldsymbol{\Theta}$ and the applied attention mask are alternately updated to refine the attention during the training.
Each stage is trained end-to-end but the two stages are trained separately since the only purpose of stage 1 is to get the initialization of the attention map for training stage 2.

\section{Experiments}
\subsection{Dataset} 
\noindent\textbf{Splenic Vascular Injury CT Dataset.} We collect 55 consecutive multi-phase CT studies with splenic vascular injury annotations, where we randomly partitioned the studies into the training and evaluation split and made sure that no patients overlap between each split, following the experimental settings in~\cite{yu2019uncertainty,zhou2019semi,xia2020uncertainty,li2020transformation,xia20203d}.
Specially, 35  cases  are  used  for  training  and  the remaining  20  cases  are  used  for  evaluation.
All patients were scanned with 40, 64, or dual source 128 CT scanners and archived at 1.5-3 mm section thickness.
Images were acquired in the arterial and portal venous phases from the dome of the diaphragm
through the greater trochanters.

\vspace{0.5ex}\noindent\textbf{Pancreatic Tumor Segmentation CT Dataset.} 
We use the 282 abdominal CT scans from the Medical Segmentation Decathlon (MSD) challenge~\cite{simpson2019large}, referred to as the pancreas tumor dataset. 
All cases are pathological cases with tumor annotation, where were collected in the portal venous phase.

\vspace{0.5ex}\noindent\textbf{Liver Tumor Segmentation CT Dataset.} 
We use the 131 contrast-enhanced 3D abdominal CT scans from the 2017 Liver Tumor Segmentation Challenge\footnote{https://competitions.codalab.org/competitions/17094\#participate-get\_data},
where the voxel-wise annotations from both liver and tumor classes are provided. The dataset was acquired by different
scanners and protocols at six clinical sites, with a largely varying in-plane resolution of 0.55 $\sim$ 1.0 mm and slice spacing of 0.45 $\sim$ 6.0 mm.
 
\vspace{0.5ex}\noindent\textbf{External Abdominal CT Dataset.} 
We use the 30 abdominal CT scans ($3779$ axial contrast-enhanced abdominal clinical CT images in total) collected under the venous phase imaging released by MICCAI 2015 Multi-Atlas Abdomen Labeling Challenge\footnote{https://www.synapse.org/\#!Synapse:syn3193805/wiki/217789} as the external set $\mathcal{S}_{ext}$.   
For each case, $13$ anatomical structures are annotated, including spleen, liver, pancreas.

\subsection{splenic vascular injury Segmentation}

\subsubsection{Technical details}
\label{sec:tech_details}
Our experiments were performed on the whole CT scan and the implementations are based on TensorFlow. All experiments were run on Titan Xp GPU.

For data pre-processing, similar to~\cite{zhou2019semi,Roth_2016_Spatial,zhu2019multi}, we simply truncated the raw intensity values to be within the range of the soft tissue CT window, \emph{i.e.}, $[-125,275]$ HU, and then normalized each raw CT case to $[0.0, 255.0]$. Random rotation of $[0,15^\circ]$ is used as an online data augmentation. 
We follow prior works~\cite{zhou2017deep,Roth_2016_Spatial} and partition all CT scans into 2D slices and our implementations are based on the widely popular 2D DeepLab-v3+ backbone\footnote{https://github.com/tensorflow/models/tree/master/research/deeplab}~\cite{chen2018encoder}.
We follow~\cite{zhou2017fixed} and only train with images which contain spleen/splenic injury. 
In the testing stage, since the groundtruth is not available, all images will be tested.
This applies to all datasets which were used in this study.
A \emph{poly} learning policy is applied with an initial learning rate of $0.08$ with a decay power of 0.9. We follow \cite{zhou2017fixed,Roth_2016_Spatial,yu2018recurrent} to use ImageNet pretrained model for initialization.  Following~\cite{chen2018deeplab}, we also apply data augmentation by randomly scaling the input images (from 1.0 to 2.5) during training. During the testing stage, multi-scale fusion is applied (scale factors are $\{1.0, 1.5, 2.0, 2.5\}$) by taking at each
position the average response across the different scales. 

We first train a segmentation network $\boldsymbol{\theta}_s$ for initial splenic attention extraction on the external set $\mathcal{S}_{ext}$. We set the number of training iteration as $50,000$ with a batch size of 16.
Then for synthetic phase augmentation, we use the official PyTorch implementation of CycleGAN\footnote{https://github.com/junyanz/pytorch-CycleGAN-and-pix2pix} and set the coefficient $\lambda=10$ in Eqn.~\eqref{Eqn:translation_losses}. We use the Adam solver
with a batch size of 2. 
The networks were trained from scratch with an initial learning rate of 0.0004 and a \emph{linear} decay policy. The whole network is trained for 50 epochs. 
Even though our collected splenic vascular injury CT dataset only contains 55 multi-phase CT, we want to emphasize that the actual number of training samples are 5,500 axial slices in each imaging phase. We find this number is sufficient to train the CycleGAN model.

The model trained on the external data $\boldsymbol{\theta}_s$ is then used to generate attention maps on both real and synthetic phases. 
We first compute the prediction for the unknown spleen class, and further expand the predicted boundary by a Gaussian filter with the standard deviation of  $\sigma=32$. To compute the initial splenic attention map, we binarize the result by using a cut-off threshold of $\rho=0.005$ in Eqn.~\eqref{eqn:init_attention}.
The splenic attention map $\mathbf{W}_\phi$ and the pseudo splenic mask $\hat{\mathbf{Y}}_{\phi, s}$ are updated for $T=2$ times.
And during each update, the real and synthetic phases, as well as the external set, are trained for $40,000$ iterations with a batch size of 16.
Unless otherwise specified, hyper-parameters $\alpha$ and $\beta$ are set as 0.5 and 0.2, respectively.

\begin{table}[t]
\renewcommand\arraystretch{1.1}
\footnotesize
\centering
\caption{Performance comparison (average DSC $\pm$ standard deviation, $\%$) on the splenic vascular injury dataset.}
\label{Tab:comparison}
\vspace{-0.6em}
\resizebox{0.7\linewidth}{!}{
\begin{tabular}{lccc}
\toprule[0.15em]
Method  & Venous & Arterial \\
\midrule
U-Net~\cite{cciccek20163d}  & 46.23 $\pm$ 19.20 & 43.12 $\pm$ 20.05 \\
V-Net~\cite{Milletari_2016_VNet}  & 43.56 $\pm$ 18.77 & 47.22 $\pm$ 19.34 \\
\midrule
C2F~\cite{zhou2017fixed}  & 47.96 $\pm$ 16.89 & 51.43 $\pm$ 17.85 \\
RSTN~\cite{yu2018recurrent}  & 50.37 $\pm$ 15.32 & 53.24 $\pm$ 16.58 \\
Attention U-Net~\cite{schlemper2019attention}  & 48.72 $\pm$ 16.55  & 52.88 $\pm$ 16.24 \\
\midrule
Ours  & \textbf{54.77 $\pm$ 13.27}  & \textbf{58.54 $\pm$ 14.90}\\
\bottomrule[0.15em]
\end{tabular}
}
\vspace{-2em}
\end{table}

\subsubsection{Evaluation Metric} 

The accuracy of segmentation is evaluated by the Dice-S{\o}rensen coefficient (DSC).
Since the injury volumes are quite different across phases, we evaluate and report the average DSC score together with the standard deviation over all testing cases are reported on each individual phase, as in~\cite{zheng2018phase}.

\begin{table*}[t]
\footnotesize
\centering
\caption{{{Ablation study on various components of our framework for multi-phase splenic vascular injury segmentation.}}}
\vspace{-0.6em}
\label{Tab:ablation}
\begin{tabular}{lccccccc}
\toprule[0.15em]
Method & $\phi=\mathrm{A}$ & $\phi=\mathrm{V}$ & $\phi=\mathrm{A'}$ & $\phi=\mathrm{V'}$ & $\phi=ext$ & Venous & Arterial \\
\midrule
\multirow{2}{*}{Separate (single-phase)}  &\checkmark &  & & & & - & 42.38 $\pm$ 22.18 \\
&  &\checkmark &  & & & 40.02 $\pm$ 21.88 & - \\
\midrule
Joint (multi-phase) &\checkmark & \checkmark & & & & 47.06 $\pm$ 17.55 & 50.05 $\pm$ 17.29 \\
\midrule
+ SynPhaseAug (w/o self-learning) &\checkmark &\checkmark  &\checkmark & \checkmark & & 50.45 $\pm$ 16.06 & 52.19 $\pm$ 16.67 \\
+ SynPhaseAug &\checkmark  & \checkmark  &\checkmark &\checkmark & & 51.36 $\pm$ 15.64 & 54.55 $\pm$ 15.10 \\
+ External Attention (w/o SynPhaseAug) &\checkmark   & \checkmark  & \checkmark & \checkmark & \checkmark & 52.89 $\pm$ 14.88  & 56.01 $\pm$ 15.14\\
+ External Attention &\checkmark   & \checkmark  & \checkmark & \checkmark & \checkmark & 54.77 $\pm$ 13.27  & 58.54 $\pm$ 14.90\\
\bottomrule[0.15em]
\end{tabular}
\vspace{-1em}
\end{table*}

\subsubsection{Results}
Performance comparison with different methods is summarized in Table~\ref{Tab:comparison} and Fig.~\ref{Fig:boxplot}(1). 
We compare our method with 2 popular medical image segmentation approaches (U-Net~\cite{Ronneberger_2015_UNet} and V-Net~\cite{Milletari_2016_VNet}) and 3 internal attention based methods (C2F~\cite{zhou2017fixed}, RSTN~\cite{yu2018recurrent}, and attention U-Net~\cite{schlemper2019attention}).
C2F and RSTN use a multi-stage framework where the first stage explicitly extracts visual attention cues, which are then fed to the second stage, whereas attention U-Net uses attention gates in the network based on non-local operators~\cite{wang2018non}.
Here ``internal attention'' refers to attention mechanisms which are established without any assistance from external data.
It is observed that, without assistance from external datasets, C2F, RSTN, and attention U-Net achieve better results on both arterial and venous phases than U-Net and V-Net, showing that attention mechanisms from internal data can already be beneficial for splenic vascular injury segmentation.
Out of these three methods, the best result is achieved by RSTN, which yields an average DSC of $50.37\%$ and $53.24\%$ on arterial and venous phases, respectively.
Meanwhile, our learning framework which leverages external splenic attention can further boost the performance to $54.77\%$ and $58.54\%$, which outperform internal attention based methods by a large margin of more than $4\%$. Our method also achieves the lowest standard deviation on both phases, demonstrating that the improvement is general and consistent.

\begin{figure*}[t]
\begin{center}
    \includegraphics[width=0.81\linewidth]{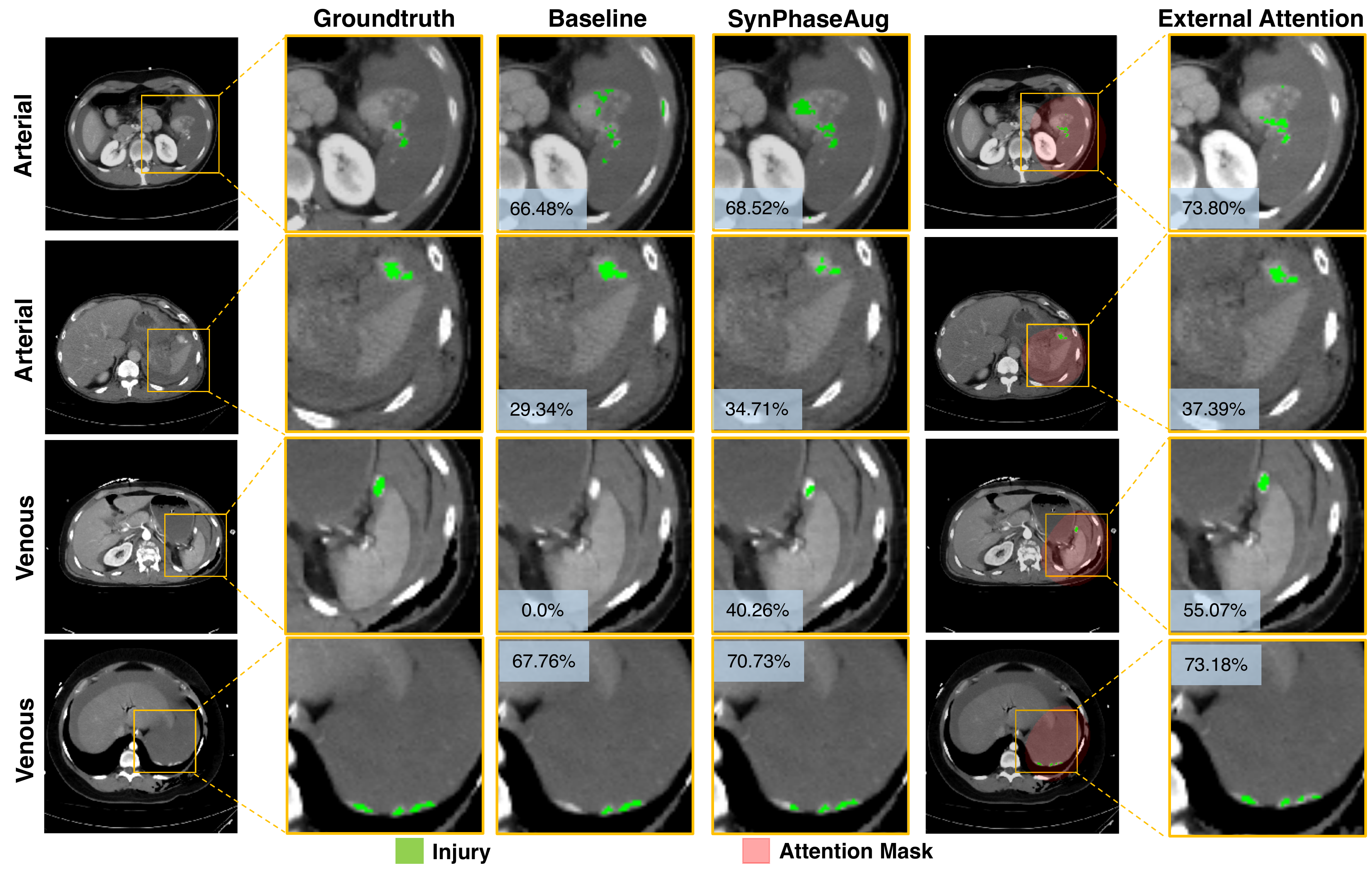}
\end{center}
\vspace{-1.5em}
\caption{Qualitative comparison. Synthetic phase augmentation already yields a performance improvement, while external attention can further enhance the segmentation accuracy. The improvement can be witnessed for both arterial and venous phases. The generated attention mask well aligns with the region-of-interest.} 
\label{Fig:qualitative}
\vspace{-1.5em}
\end{figure*}

\vspace{-1em}
\subsection{Ablation Study}
We conduct ablations to analyze the influence of different designs, components and hyper-parameters in our approach. 
\subsubsection{On the effectiveness of synthetic phase augmentation}
We perform experiments by varying the input imaging phase $\phi$~from~$\{\mathrm{A,V,A',V'},ext\}$. 
Here we define \emph{Joint} as directly training on the union of both imaging phases.
And the trained model is then used for evaluating results on both arterial and venous phases.
\emph{Separate} refers to the setting where only single-phase images are available, therefore the training and evaluation is performed separately on either the arterial or the venous phase.
From Table~\ref{Tab:ablation}, we can observe that \emph{Joint} outperforms \emph{Separate} significantly, due to the involvement of both phases in the learning process.

Built upon \emph{Joint}, the model performance can be further enhanced when training with additional synthetic phase images as augmentation (indicated by \emph{SynPhaseAug} in Table~\ref{Tab:ablation}). 
Here, we note our segmentation network is jointly trained on both the real and the synthetic data in a self-learning manner, \emph{i.e.}, we use pseudo-labels of the injury class on synthetic phases $\mathrm{A'/V'}$ computed by the \emph{Joint} model for training rather than the annotated labels from the original real phases $\mathrm{A/V}$.
As mentioned in Sec.~\ref{sec:multi-phase}, our intuition lies in that a number of fake injury regions can be inpainted on the generated images (see examples in Fig.~\ref{Fig:gan_example}), which makes the usage of pseudo-labels more reasonable than the original labels.

Our experimental results also validate this intuition, \emph{i.e.}, synthetic phase augmentation without self-learning (indicated by \emph{SynPhaseAug (w/o self-learning)} in Table~\ref{Tab:ablation}) results in performance degradation as large as $2.36\%$ on the arterial phase in segmentation accuracy.
With self-learning, the performance improvement compared with \emph{Joint} can be more than $4\%$ in terms of average DSC, boosting the performance on the venous and the arterial phase from $47.06\%$ and $50.05\%$ to $51.36\%$ and $54.55\%$ respectively. 
We have also provided the performance comparison in boxplot (see Fig.~\ref{Fig:boxplot}(2)).
This large performance improvement suggests that the joint training with synthetic phase augmentation in a self-learning manner not only enables an augmented training set, but also allows a more thorough knowledge integration from both phases. The benefits of synthetic phase augmentation are also demonstrated in Fig.~\ref{Fig:qualitative}.

\subsubsection{On the effectiveness of external attention}
For multi-phase splenic vascular injury segmentation, our external attention mechanism is built upon synthetic phase augmentation by default. All real and generated images form our internal set whereas an external set of abdominal CT scans is used for extracting splenic attention.
We also use CycleGAN to train a domain translation model between the internal and the external set, for the purpose of augmentation. All hyper-parameter settings are the same as those used for synthetic phase augmentation.

\begin{figure}[t]
\begin{center}
    \includegraphics[width=\linewidth]{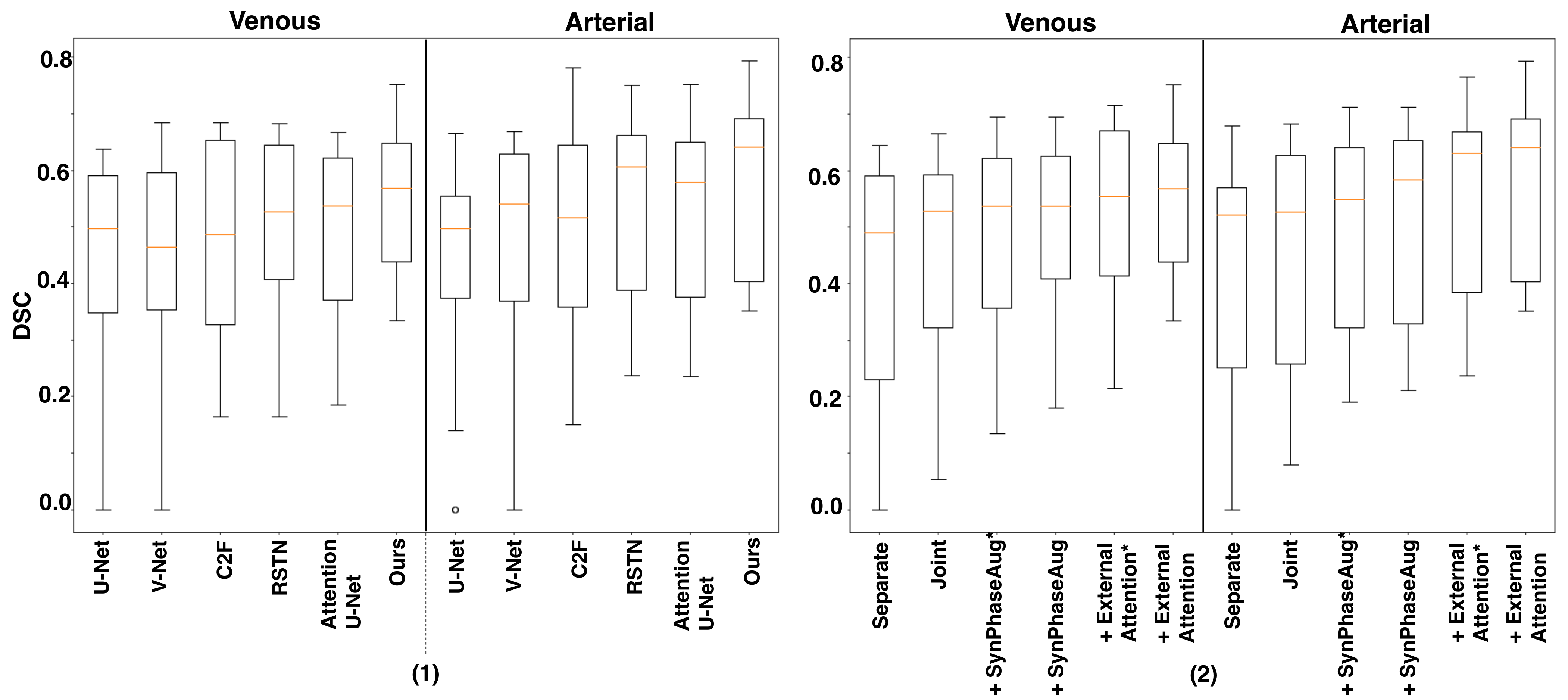}
\end{center}
\vspace{-1.5em}
\caption{ Performance comparison (DSC) in box plots of multi-phase splenic injury segmentation: (1) Comparison with different segmentation methods. Our method ourperforms all other competitors in both Venous and Arterial phases. (2) Ablation study on various components of our framework. Both synthetic phase augmentation and external attention are crucial for improving the model performance. For simplicity, SynPhaseAug* refers to SynPhaseAug (w/o self-learning) and External Attention* refers to External Attention (w/o SynPhaseAug). }
\vspace{-1em}
\label{Fig:boxplot}
\end{figure}

\begin{figure}[t]
\begin{center}
    \includegraphics[width=0.9\linewidth]{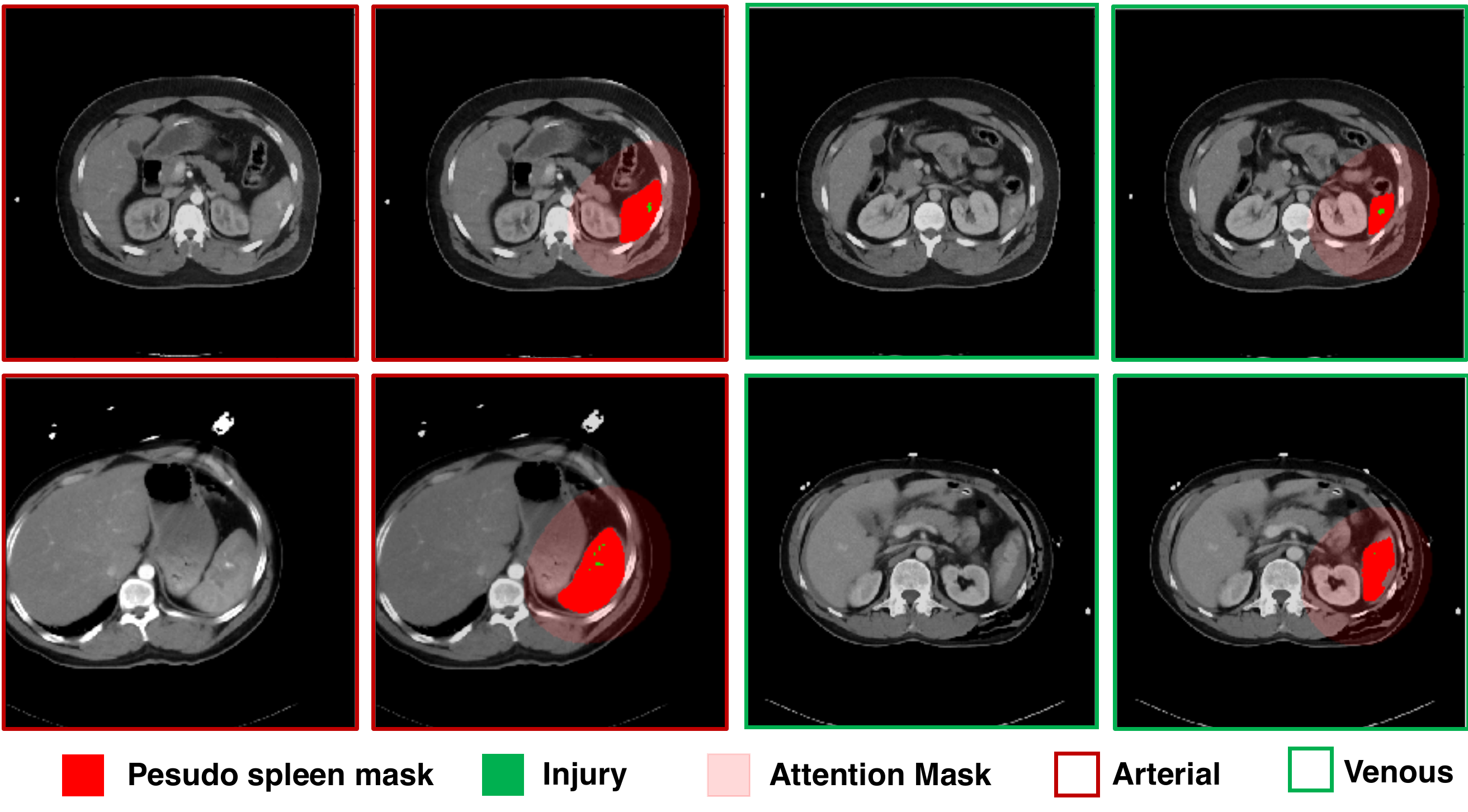}
\end{center}
\vspace{-1.5em}
\caption{Qualitative examples of refined pseudo spleen masks as well as the expanded attention masks.} 
\vspace{-1em}
\label{Fig:spleen_mask}

\end{figure}

After initial attention map generation, the attention map, pseudo-labels as well as network weights are iteratively updated. Throughout the iterations, we find that the quality of the pseudo splenic mask can be improved. 
To quantify this, we calculate the recall rate as the ratio between the number of retrieved injury voxels and the number of all injury voxels. The recall rates are improved from 94.55\% and 96.37\% to 98.96\% and 99.79\% on the venous and the arterial phase respectively.
We also show a set of qualitative examples of refined pseudo spleen masks as well as the attention masks in Fig.~\ref{Fig:spleen_mask}, which demonstrates the high quality of the refined masks.

As aforementioned, compared to \emph{Joint}, \emph{i.e.}, directly training with multi-phase images, training with additional synthetic phases (\emph{SynPhaseAug}) already suggests large benefits.
Nevertheless, from Table~\ref{Tab:ablation}, we observe exploiting external attention can further lead to a performance gain of $3.41\%$ and $3.99\%$ on each phase compared with \emph{SynPhaseAug}.
The performance comparison in boxplot (see Fig.~\ref{Fig:boxplot}(2)) further suggests the significance of the improvement.
We also demonstrate the effectiveness of the proposed external attention in a set of qualitative examples, \emph{e.g.}, as shown in Fig.~\ref{Fig:qualitative}, the computed attention masks are well aligned with the region-of-interest, which effectively reduces the irrelevant responses from the background region during the learning process.
Last but not the least, it is also worth mentioning that even without synthetic phase augmentation, pure external attention (indicated by \emph{External Attention (w/o SynPhaseAug)} in Table~\ref{Tab:ablation}) also achieves superior results than \emph{Joint} by a large margin of $5.83\%$ and $5.96\%$ in the arterial and venous phase, respectively.
The generalization of pure attention is also well justified in Sec.~\ref{sec:liverTumor} and Sec.~\ref{sec:PanTumor}.

\subsubsection{Diagnosis on hyper-parameters $\alpha$ and $\beta$}
\begin{figure}[t]
\begin{center}
    \includegraphics[width=\linewidth]{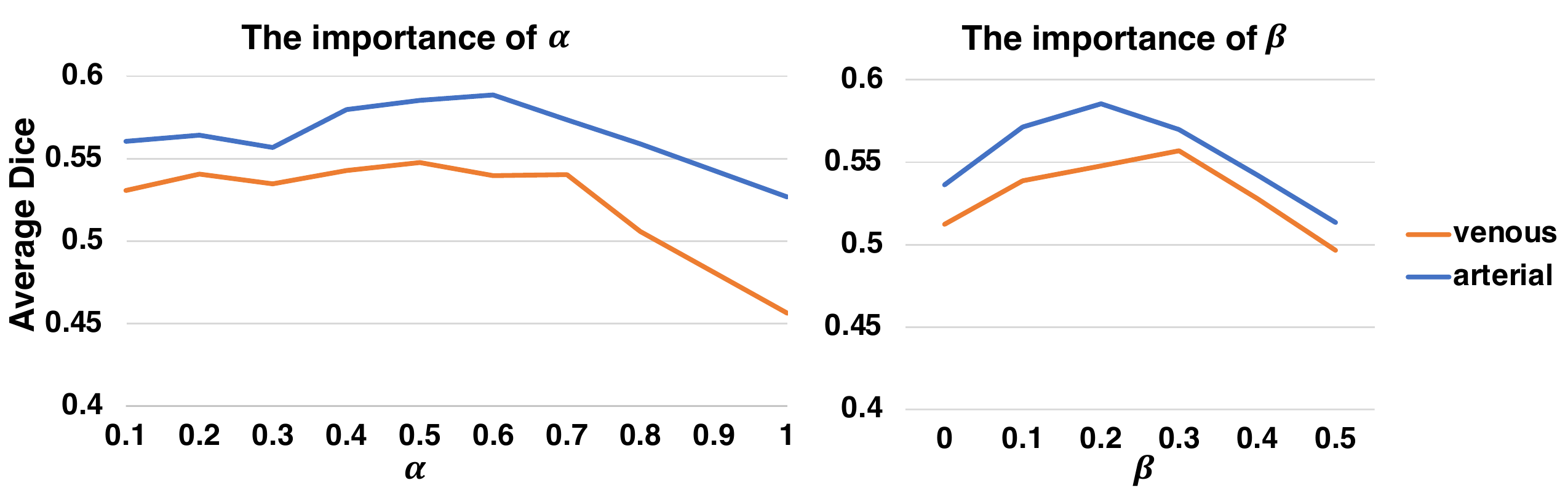}
\end{center}
\vspace{-1.0em}
\caption{The importance of hyper-parameters $\alpha$ and $\beta$.} 
\label{Fig:hyper-parameter}
\vspace{-1.0em}
\end{figure}

$\alpha$ is used for balancing the weight between the real and synthetic data, whereas $\beta$ is used for adjusting the weight between internal and external data during the joint training.
Here we provide an
ablation analysis to study the importance of $\alpha$ and $\beta$ in Eqn.~\eqref{Eqn:total_seg_loss}. By varying
the value of $\alpha$ and $\beta$ from 0.0 - 1.0 and 0.0 - 0.5 respectively, the results are summarized 
in Fig.~\ref{Fig:hyper-parameter}. 
We first fix $\beta = 0.2$ and show how different choices of $\alpha$ can affect the overall performance. 
From the results, we can see that $\alpha = 0.5$ achieves the best result, which suggests that features from real and synthetic data should be equally learned to yield the best model.
Meanwhile, we also note that we should avoid making the portion of synthetic imaging data over 0.7 ($\alpha > 0.7$) during training.
Drawing $\alpha$ near 0.5 yields the most robust segmentation results on both testing phases.
On the contrary, $\alpha = 0.0$ (\emph{i.e.}, only real images are used for training) or $\alpha = 1.0$ (\emph{i.e.}, only synthetic images are used for training) achieves the worst results.

We then fix $\alpha = 0.5$ and show how different choices of $\beta$ will affect the overall performance.
We observe that $\beta = 0.2$ achieves the best
performance. Meanwhile, we note that our approach is not sensitive
to this hyper-parameter when $\beta$ is drawn between 0.1-0.3. 

\begin{figure}[t]
\begin{center}
    \includegraphics[width=0.8\linewidth]{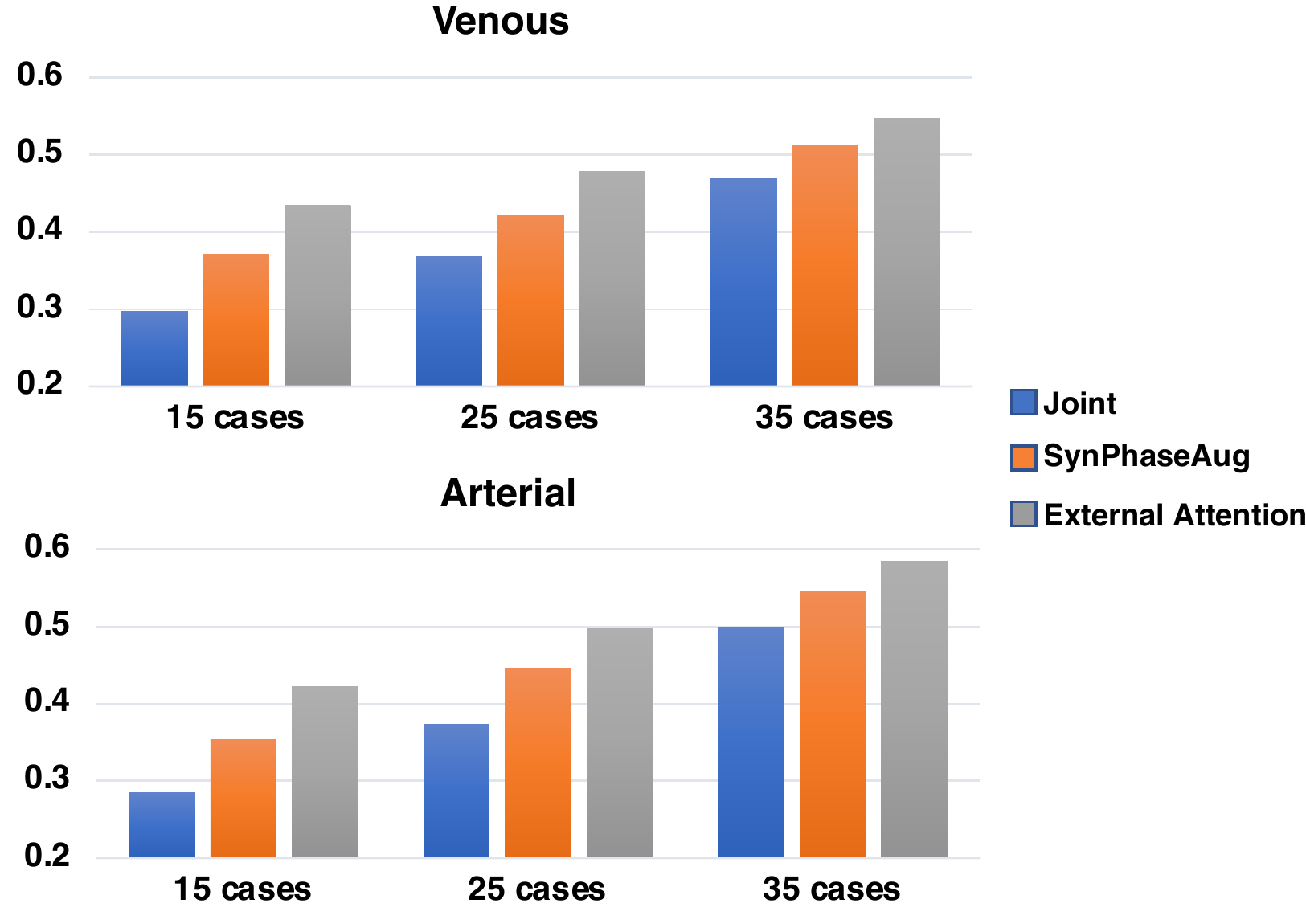}
\end{center}
\vspace{-1.0em}
\caption{Ablation study on the number of training cases.} 
\label{Fig:less_supervision}
\vspace{-1em}
\end{figure}

\subsubsection{The number of training cases}
We also reduce the number of training cases from 35 to 15 and 25 to see the effects of synthetic phase augmentation and external attention with even less supervision.
The results are summarized in Fig.~\ref{Fig:less_supervision}. Both \emph{SynPhaseAug} and \emph{External Attention} leads to larger performance gain with fewer training cases. 

This suggests that both synthetic phase augmentation and external attention are beneficial for multi-phase splenic vascular injury segmentation, especially with limited data.

\subsubsection{The number of iteration $T$}
Next, we show how the segmentation results vary with different iteration number $T$ in Table~\ref{tab:iteration}. $T = 0$ means that no external attention is applied. We can see that at first the performance increases when $T$ goes up, and then begins to saturates after $T$ reaches 2.

\begin{table}[tbp]
\footnotesize
\centering
\caption{Ablation study on the number of iteration $T$.}
\label{tab:iteration}
\vspace{-0.5em}
\begin{tabular}{lcc}
\toprule[0.15em]
$T$ & Venous & Arterial \\
\hline
0 & 51.36 $\pm$ 15.64 & 54.55 $\pm$ 15.10  \\
1 & 54.15 $\pm$ 14.13 & 57.08 $\pm$ 15.01  \\
2 & 54.77 $\pm$ 13.27  & 58.54 $\pm$ 14.90  \\
3 & 55.05 $\pm$ 13.29  & 57.96 $\pm$ 14.77  \\
4 & 54.29 $\pm$ 13.53  & 58.70 $\pm$ 14.85  \\
\bottomrule[0.15em]
\end{tabular}
\vspace{-2em}
\end{table} 

\subsection{Generalization to Liver Tumor Segmentation}
\label{sec:liverTumor}
Following \cite{xia20203d}, we also report the average DSC on the 131 abdominal CT scans of the liver tumor segmentation CT dataset, with a random split of 100 training cases and 31 cases for validation.
We also test on smaller training sets, \emph{i.e.}, training with only 10 cases and 20 cases (indicated by \emph{10\% supervised} and \emph{20\% supervised}), to show the generalization of our approach on liver tumor segmentation, especially with limited training data. 
Similar to previous settings, only the annotation of the liver tumor is utilized in our training. 
Specifically, for our approach (indicated by ``ours''), the 30 cases of the external abdominal CT dataset $\mathcal{S}_{ext}$ along with the liver supervision are used for extracting liver attention. 
Here, the external attention is instantiated without applying synthetic phase augmentation (\emph{i.e.}, $\alpha$ is set as 1.0 in Eqn.~\eqref{Eqn:total_seg_loss}) since multi-phase information is not accessible for the liver tumor dataset (\emph{i.e.}, only venous phase images are provided).
All other implementation details are the same as described in Sec.~\ref{sec:tech_details}.

The performance comparison is summarized in Table~\ref{tab:LiverTumor}, where we can see that our approach achieves competitive results compared with existing methods, including V-Net~\cite{Milletari_2016_VNet}, 3D ResNet-18~\cite{xia20203d}, U-Net~\cite{cciccek20163d}, nn-UNet~\cite{isensee2021nnu}, DeepLab V3+~\cite{chen2018encoder}. It is also noteworthy to mention that by fully exploiting external attention, our approach demonstrates greater superiority with less supervision. For instance, our method outperforms 3D ResNet-18 and U-Net by a large margin of $3.63\%$ and $9.68\%$ under 20\% supervision. 
Under 10\% supervision, our performance surpasses those of ResNet-18 and U-Net by even further of $4.05\%$ and $11.99\%$.
Compared with the state-of-the-art nn-UNet, our method obtains similar results under 100\% supervision, and outperforms it by $2.06\%$ and $2.88\%$ under 10\% and 20\% supervision.
This promising result suggests the generalization of the proposed external attention.

\vspace{-0.5em}
\subsection{Generalization to Pancreatic Tumor Segmentation}
\label{sec:PanTumor}
We also report the segmentation results on the pancreas tumor segmentation dataset. Following the evaluation protocol in~\cite{xia2020uncertainty}, we partition the whole dataset into 200 cases for training and 81 cases for validation. In the training stage, we use the tumor annotation as supervision; in the testing stage, our goal is to segment the tumor region from the whole CT.
Other settings are similar to the liver tumor segmentation experimental settings, \emph{i.e.}, we use the 30 cases of the external abdominal CT dataset $\mathcal{S}_{ext}$ along with the pancreas supervision for extracting pancreas attention for our approach (denoted as ``ours''). 
Again, synthetic data augmentation is not applicable here since only single-phase images are provided in the pancreas tumor dataset. We compare our method with ResDSN~\cite{zhu2019multi}, the state-of-the-art architecture on pancreatic ductal adenocarcinoma detection, and V-Net~\cite{Milletari_2016_VNet}, nn-UNet~\cite{isensee2021nnu},
DeepLab V3+~\cite{chen2018encoder}.
We also evaluate our approach with 20 and 40 training cases (\emph{i.e.}, \emph{10\% supervised} and \emph{20\% supervised}), to demonstrate the effectiveness of external attention with limited supervision.

The performance comparison is summarized in Table~\ref{tab:PancreasTumor}. Our method achieves superior performance under all settings. Notably, when fewer training cases are provided, the proposed external attention demonstrates greater significance compared with prior arts. For instance, our segmentation accuracy is $4.61\%$ and $6.6\%$ higher than ResDSN with 20\% and 10\% supervision respectively.

\begin{figure}[tbp]
\begin{center}
    \includegraphics[width=\linewidth]{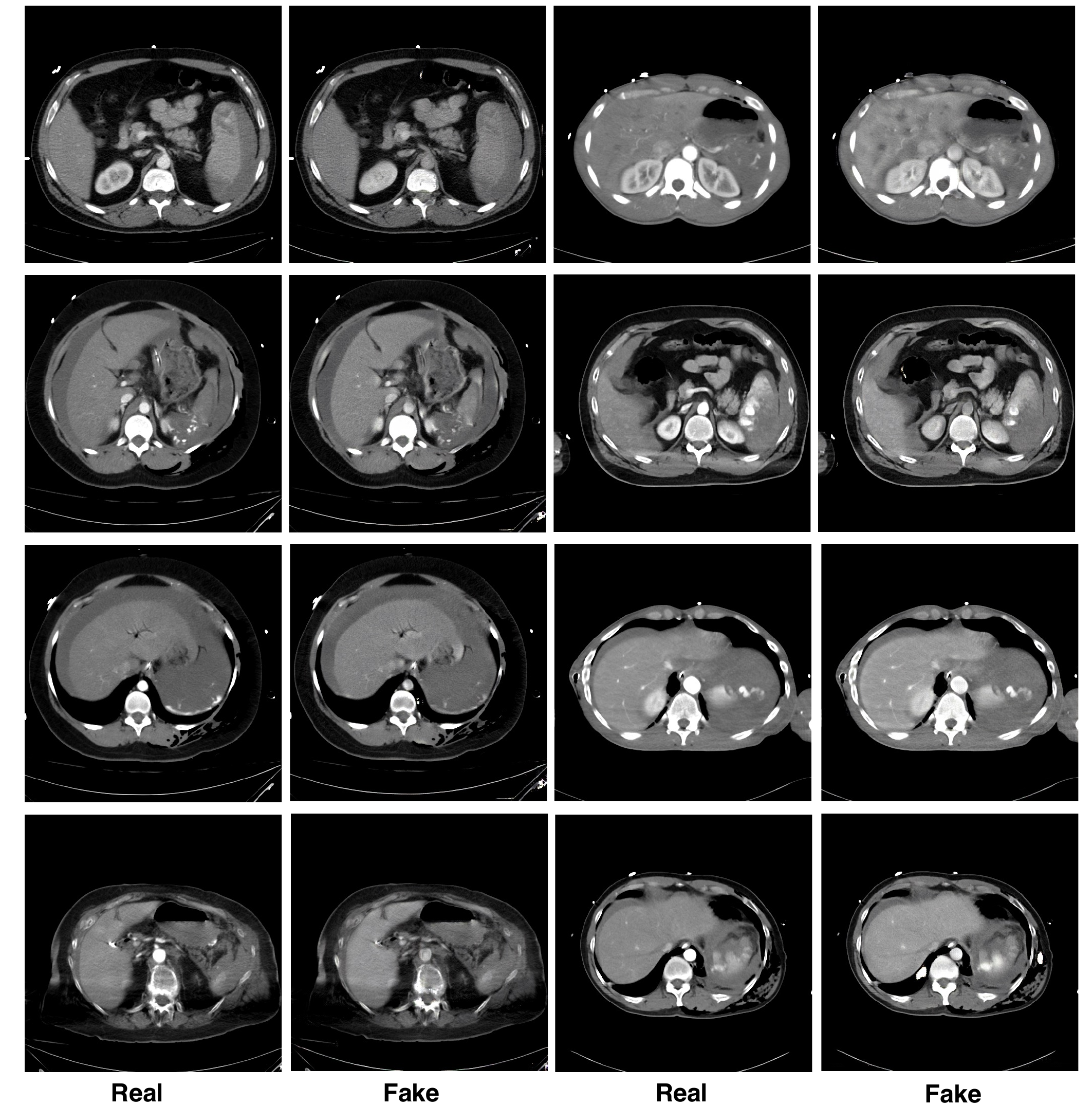}
\end{center}
\caption{Qualitative examples of generated synthetic phase images.} 
\label{Fig:more_gan_results}
\end{figure}

\begin{table}[t]
\footnotesize
\centering
\caption{{Comparison (DSC, \%) on the liver tumor dataset.}}
\label{tab:LiverTumor}
\vspace{-0.5em}
\begin{tabular}{lccc}
\toprule[0.15em]
\multirow{2}{*}{Method} & 100\%  & 20\%  & 10\%  \\
 &supervised &supervised &supervised \\
\midrule
V-Net~\cite{Milletari_2016_VNet}   &63.02  &48.40 &41.92   \\
3D ResNet-18 (MV)~\cite{xia20203d}  &65.65  &53.15 &48.90   \\
3D ResNet-18 (Single)~\cite{xia20203d}  &64.00  &50.39 &43.98   \\
U-Net~\cite{cciccek20163d} &62.24 &47.10 &40.96 \\
nn-UNet~\cite{isensee2021nnu} &65.78 &54.72 &50.07\\
DeepLab V3+~\cite{chen2018encoder} &64.86 &51.58 &45.62\\
Ours &65.98 &56.78 &52.95 \\
\bottomrule[0.15em]
\end{tabular}
\vspace{-1.5em}
\end{table}

\begin{table}[t!]
\footnotesize
\centering
\caption{{Comparison (DSC, \%) on the pancreas tumor dataset.}}
\label{tab:PancreasTumor}
\vspace{-0.5em}
\begin{tabular}{lccc}
\toprule[0.15em]
\multirow{2}{*}{Method} & 100\%  & 20\%  & 10\%  \\
 &supervised &supervised &supervised \\
\midrule
V-Net~\cite{Milletari_2016_VNet} &47.53 &38.97 & 35.68 \\
ResDSN~\cite{zhu2019multi}  &49.22  &41.72 &38.25   \\
nn-UNet~\cite{isensee2021nnu} &51.03 &44.12 & 42.31 \\
DeepLab V3+~\cite{chen2018encoder} &49.96  &42.38 &39.52 \\
Ours &51.63 &46.33 &44.85 \\
\bottomrule[0.15em]
\end{tabular}
\vspace{-2em}
\end{table}

\section{Discussion}
We have presented a novel approach for splenic vascular injury segmentation from multi-phase trauma CT scans.
Extensive experiments are conducted on our curated dataset and show the effectiveness of our method.
To the best of our knowledge, this is the first study that addresses this important clinical use case with deep learning to-date. 
We hope our study can inspire more researchers to develop machine learning techniques for trauma radiology, which has not drawn much research attention to-date.

In our work, we propose to exploit external data for iterative splenic attention refinement, which can be served as important guidance for splenic vascular injury detection, via mining the unknown class. 
This principle can be widely applied to different tasks. 
For instance, we have also demonstrated the effectiveness of using additional abdominal CT with pancreas/liver supervision to facilitate the relatively harder tumor segmentation. 
We have also shown more results generated from CycleGAN in Fig.~\ref{Fig:more_gan_results} to further demonstrate the effectiveness of our proposed synthetic phase augmentation module.
We hope our method can offer insights for future research regarding detecting affiliated lesions of an organ.

There are several limitations for our proposed work. For one, the current study does not consider the segmentation of multi-phase splenic injury from multi-site CT yet.
Future work should also increase the scale of the current study as well as conducting further characterization of the injury type.

\section{Conclusion}
This paper addresses the problem of multi-phase splenic vascular injury segmentation, especially with limited data. 
Specifically, we propose to mine external attention to enhance the associated injury segmentation. 
We additionally introduce synthetic phase augmentation for populating the training set, which further enhances the learning process from multi-phase images.
Extensive experiments have shown the effectiveness of both external attention and synthetic phase augmentation, especially with fewer training cases.
The generalization of our approach is further validated on both liver tumor segmentation and pancreas tumor segmentation.

\bibliographystyle{IEEEtran}

\bibliography{bare_jrnl}

\end{document}